\documentclass[aps, prd, amsmath, a4paper, 10pt, twocolumn, nofootinbib, showkeys]{revtex4-2}
\usepackage[english]{babel}
\usepackage{amssymb}

\usepackage{graphicx}
\usepackage[colorlinks=true, allcolors=blue]{hyperref}
\usepackage{enumitem}

\newcommand{\Det}{\text{Det}\,}

\begin{document}
\title{Cosmological phase transitions from the functional measure}
\author{Bruno \surname{Berganholi}$^1$}
\author{Gl\'auber C. \surname{Dorsch}$^1$}
\email{glauber@fisica.ufmg.br}
\author{Iber\^e Kuntz$^2$}
\author{Beatriz M. D. \surname{Sena}$^1$}
\author{Giovanna F. do Valle$^1$}
\affiliation{\vskip2mm $^1$Departamento de F\'isica,  Universidade Federal de Minas Gerais (UFMG), Belo Horizonte, MG, Brazil\\
$^2$Departamento de F\'isica, 
Universidade Federal do Paran\'a (UFPR),
Curitiba, PR, Brazil}

\begin{abstract}
    We investigate how competitive gravitational wave detectors can be to current and near-future colliders in probing a model where the new physics is completely encapsulated in a modified scalar sector. For this, we study a model where an additional logarithmic term arises in the scalar potential due to a non-trivial path integral measure, which is constructed using effective field theory. This new term alters the dynamics of cosmological phase transitions and could lead to potentially detectable gravitational waves from the early Universe. Our results confirm the expectation that the intensity of such spectrum is highly correlated to the scalar field's self-coupling, and that gravitational wave experiments could therefore be used to probe these couplings with an accuracy competitive with the projected sensitivity of near-future colliders.
\end{abstract}
\maketitle

\section{Introduction}

Since the dawn of high energy physics, particle accelerators have continuously reinforced their status as \emph{the} paradigmatic tools for testing our models of fundamental particle interactions. However, the initial success of the LHC program in detecting the Higgs boson has so far been followed by a disturbing silence regarding concrete signs of new physics, despite prior expectations to the contrary. At the same time, the last decade has been marked by the inaugural direct detection of gravitational wave (GW) signals, even coming from experiments employing considerably different detection techniques, such as interferometers~\cite{LIGOScientific:2016aoc, LIGOScientific:2018mvr, LIGOScientific:2021usb, KAGRA:2021vkt} and pulsar timing arrays~\cite{NANOGrav:2023gor, EPTA:2023fyk}. These experiments open a new window for exploring our cosmos, and could even be employed as new probes for testing particle physics models. Most importantly, they could bring new information on the high energy frontier, competitive with or even complementary to those coming from colliders~\cite{Arcadi:2023lwc}. This is especially the case for models where the new physics affects mostly (or exclusively) the scalar sector, since probing the details of the Higgs potential involves measuring its self-couplings, which is a formidable task for current colliders~\cite{CMS:2022dwd}, and challenging even at accelerators projected for the near-future~\cite{DiVita:2017eyz, DiVita:2017vrr, Mangano:2020sao}.

An interesting example of these types of models can be found in the recent literature. It has been noted that the definition of the functional measure in the partition function of quantum field theories largely depends on the geometry of configuration space, and adopting the Riemannian measure amounts to effectively adding a logarithmic correction to the scalar potential~\cite{Kuntz:2022kcw, Kuntz:2024opj}. This new term leads to a correction of the Higgs quartic coupling as well as to $\mathcal{O}(\phi^6/\Lambda^2)$ operators (with $\Lambda$ some cutoff scale). 
Recent works have shown that this effect has no impact on electroweak precision observables~\cite{Kuntz:2024opj}, so the most immediate direct probe at colliders would be via the modified Higgs trilinear couplings\footnote{Directly probing the quartic couplings or deviations from an effective $\phi^6$ operator would be rather hopeless in near-future colliders, as argued above.}. At the same time, such a modification of the scalar potential could have a drastic impact on the cosmic evolution of the Universe, even altering the nature of the electroweak phase transition, and leading to a potentially detectable GW spectrum in near-future interferometers. 

The purpose of this work is to investigate the cosmological impact of a non-trivial Riemannian functional measure, and explore the possibility of probing these effects with GW detectors. We will see that future interferometers, such as LISA, DECIGO and BBO, could be sensitive to new logarithmic terms in the effective potential. This can be translated into bounds on the trilinear coupling, and we will show that future GW experiments could probe these at a competitive level if compared to the projected sensitivity of near-future experiments. 

This is important because we are living in an era where a number of GW experiments are under the spotlight, and many others are receiving increasing attention and funding, so it is all the more important that we harness the capabilities of these machines to probe particle physics as well. In a recent work we have illustrated how direct particle detectors and GW interferometers could provide \emph{complementary} information on models with dark sectors~\cite{Arcadi:2023lwc}. Here we show how \emph{competitive} these GW experiments can be as compared to near-future colliders for probing the Higgs self-couplings if the new physics is exclusively in the scalar sector. However, it is important to emphasize at the outset that investing on both fronts (i.e. GW detectors and colliders) will remain essential for the progress of particle physics in the near future, especially at this moment when we still know little about the actual character of this BSM physics.

The paper is organized as follows. In section~\ref{sec:measure} we review the issue of defining an integration measure in field space and show that a redefinition of this quantity in the partition function amounts to adding a logarithmic correction to the effective scalar potential. Section~\ref{sec:physicality} is dedicated to discussing the physicality conditions one can impose on the model. In section~\ref{sec:Vth} we introduce the thermal corrections and discuss the dynamics of the phase transition and its relevant parameters, which leads to gravitational wave production and its potential detection at future interferometers such as LISA, DECIGO and BBO, discussed in section~\ref{sec:GW}. Our results are shown in section~\ref{sec:results}, and our conclusions are reserved for section~\ref{sec:conclusions}.

\section{Functional measure in Quantum Field Theory}
\label{sec:measure}

Path integrals have long become part of the arsenal of every field theorist, with applications ranging from condensed matter to high-energy physics. The functional approach shines, in particular, when applied to gauge theories as it preserves manifested Lorentz covariance and gauge invariance. Path integrals are also a centerpiece in Wilson's modern formulation of effective field theory, which could be taken as the proper definition of quantum field theory \cite{Wilson:1983xri,KCostello}.

In Wilson's approach, the theory is finite and defined at energies $\Lambda < \Lambda_\text{phys}$ below some physical scale $\Lambda_\text{phys}$~\footnote{We are adopting DeWitt's index convention by merging continuous (spacetime coordinates) and discrete indices (denoted by capital letters) into small indices $i = (x, I)$.
For example, if $\varphi^i = A^a_\mu(x)$ is a vector field, $i = (x, \{\mu, a\})$. Repeated small indices then implicitly contain sums over discrete indices and integrations over spacetime.}: 
\begin{equation}
	Z_\Lambda[J]
	=
	\int_{\Omega(\Lambda)}\mathrm{d}\mu[\varphi] 
	\, e^{- \left( S[\varphi] + J_i \varphi^i \right)},
	\label{Z}
\end{equation}
where $S[\varphi]$ is the Euclidean classical action for some arbitrary field $\varphi^i$
and $J_i$ is the external current. Here $\Omega(\Lambda)$ is the integration domain that includes modes with energy below $\Lambda$. The physical scale $\Lambda_\text{phys}$, such as the interatomic scale in condensed matter or the Planck length in high-energy physics, acts as a true regulator at high energies. Unlike the ordinary approach to quantum field theory, $\Lambda_\text{phys}$ prevents UV divergences from appearing, making the theory finite from the onset. Renormalization in this context is just a statement about the independence of observables under changes in $\Lambda$:
\begin{equation}
	\Lambda \frac{d Z_\Lambda[J]}{d \Lambda} = 0\, .
	\label{RG}
\end{equation}

Although Wilson's effective field theory resolves some conceptual issues regarding the definition of quantum field theory (it is finite after all), the integration measure $\mathrm{d}\mu[\varphi]$ remains an obscure object \cite{Unz:1985wq,Toms:1986sh,Moretti:1997qn,Hatsuda:1989qy,vanNieuwenhuizen:1989dx,Armendariz-Picon:2014xda,Becker:2020mjl,Buchbinder:1987vp,Hamamoto:2000ab}. The majority of the literature just take this measure to be trivial:
\begin{equation}
	\mathcal{D}\varphi^i
	=
	\prod_i \mathrm{d}\varphi^i
	\, .
	\label{trivialm}
\end{equation}
The choice of the integration measure, however, depends on geometrical and topological aspects of the configuration space. Even if one takes the natural choice, namely the measure induced by the underlying topology, it is unlikely that Eq.~\eqref{trivialm} will suffice in general. For one, fields can be singular (either diverge or not be defined) at some regions, which already shows that the configuration space is not simply connected. The presence of constraints, such as in gauge theories, would also make the configuration-space topology non-trivial, preventing the use of the trivial measure \eqref{trivialm}.
These non-trivial characteristics of the configuration space are inherited from the phase space structure implied by the canonical quantization (see the Appendix of \cite{Kuntz:2024opj}).

From the theory of integration on manifolds, albeit we are working in the quite unusual setting of an infinite-dimensional manifold \cite{DeWitt:2003pm}, the correct measure ought to be
\begin{equation}
    \mathrm{d}\mu[\varphi] = \mathcal{D}\varphi^i \sqrt{\Det G_{ij}}
    \ .
    \label{measure}
\end{equation}
Here $\Det G_{ij}$ denotes the functional determinant, which includes the determinant of finite-dimensional operators (hereby denoted by $\det$), of the  configuration-space metric $G_{ij}$.
One should recall from differential geometry that the metric $g$ is an additional structure independent of the underlying topological manifold $\mathcal{M}$. There are many possible metrics for the same $\mathcal M$, which results in different Riemannian manifolds $(\mathcal M, g)$ for every choice of $g$. Incidentaly, one faces the problem of choosing $G_{ij}$. 

Typically, the configuration-space metric is identified from the action's kinetic term, as in non-linear sigma models \cite{Finn:2019aip, Vilkovisky:1984st, Fradkin:1976xa, Fradkin:1973wke}. In this approach, the measure can be used to cancel out ultralocal divergences \cite{Fradkin:1976xa, Fradkin:1973wke}. The function $G_{IJ}$ are then defined from the kinetic term coefficients, namely:
\begin{align}
	&\int \mathrm{d}^4x \,
	G_{IJ}(\varphi) \, \partial_\mu\varphi^I \partial^\mu \varphi^J
        \\
	&=
	\int \mathrm{d}^4x \,
	\int \mathrm{d}^4x' \,
	G_{IJ}(\varphi) \, \delta^{(4)}(x-x')\, \frac{\partial \varphi^I(x)}{\partial x^\mu} \frac{\partial \varphi^J(x')}{\partial x'_\mu}
	\ ,
        \nonumber
\end{align}
hence~\footnote{The configuration space $\mathcal C$ is the set of all fields $\varphi^I(x)$. The set $\mathcal N_{x} \subset \mathcal C$ defined by all configurations at a fixed point $x^\mu$ is a finite-dimensional manifold. The function $G_{IJ}$ is the metric over $\mathcal N$.}
\begin{equation}
	G_{ij}
	=
	G_{IJ}(\varphi)
	\, \delta^{(4)}(x-x')
	\ .
	\label{ultralocal}
\end{equation}
This proportionality with Dirac's delta is referred to as ultralocality and follows from the assumption of a local Lagrangian. This definition via the kinetic term can be justified when the phase-space measure
\begin{equation}
	\mathrm{d}\mu[\varphi,\Pi] = \mathcal{M}(\varphi,\Pi) \mathcal{D}\varphi^i \mathcal{D} \Pi_i
\end{equation}
is trivial, i.e. $\mathcal{M}(\varphi,\Pi) \equiv 1$.
In this case, the determinant factor in Eq.~\eqref{measure} follows upon integration over the canonical momentum $\Pi_i$. This argument also relies on the Gaussian integration, thus it cannot be applied to higher-derivative theories where higher powers of $\Pi_i$ is present in the Hamiltonian.

Even when the measure is admittedly non-trivial, the standard practice appeals to dimensional regularization to trivialize it. The usual rationale goes by noting that
\begin{align}
    \Det G_{ij}
    &=
    \exp\left(\delta^{(n)}(0) \int\mathrm{d}^nx \, \log\det G_{IJ}\right)
    \ ,
    \label{dimr}
\end{align}
for any ultralocal metric of the form \eqref{ultralocal},
and that scaleless integrals vanish in dimensional regularization, hence $\delta^{(n)}(0)=0$ and $\Det G_{ij} = 1$. This is an improper use of dimensional regularization because the argument of the exponential in Eq.~\eqref{dimr} is the product of formally divergent quantities~\cite{DeWitt:2003pm}. The integral in Eq.~\eqref{dimr} can be divergent too,
thus it must be carefully handled, together with the Dirac delta, before simply setting the exponential to unity. That is precisely the sort of undefined products one encounters when anomalies are present, in which case a naive use of dimensional regularization would erroneously hide the violation of classical symmetries \cite{Fujikawa:1979ay, Fujikawa:1980eg, Fujikawa:1980vr}. We shall give another example of this later on.

We should stress that there is no universal choice for $G_{ij}$, which must be regarded as part of the theory's definition. Different metrics on the configuration space correspond to distinct quantization schemes. 
The complete specification of the theory is therefore given by the pair $(S[\varphi], G_{ij})$. Choosing different metrics for the same classical action leads to distinct quantum theories derived from the same classical framework.

The aforementioned approach, whereby one defines $G_{ij}$ from the kinetic terms \cite{Finn:2019aip, Vilkovisky:1984st, Fradkin:1976xa, Fradkin:1973wke}, is natural as the metric is induced by the most fundamental physical object, thus requiring only $S[\varphi]$ to fully specify the quantum theory. However, as we argued, this definition faces a number of limitations for not being applicable to higher-derivative theories or non-trivial phase spaces.
Ref.~\cite{Kuntz:2024opj} introduced a novel approach where
the configuration-space metric $G_{ij}$ is constructed using effective field theory, writing it as an infinite tower of invariant operators. 
From this perspective, the definition of $G_{ij}$ is decoupled from the classical action $S[\varphi]$, but the same symmetry principles of effective field theory are adopted to define them both independently.
This definition provides a reliable description at energies below the physical cutoff $\Lambda_\text{phys}$, irrespective of the exact functional measure required in a UV-complete theory. In this framework, the free parameters in $G_{ij}$ influence the theory's phenomenology and can be determined through experimental observations \cite{Kuntz:2022kcw, Kuntz:2024opj, deFreitas:2023ujo, Casadio:2024vfh} rather than solely by their relationship to the kinetic term.

In this effective field theory approach, it is still reasonable to assume ultralocality as in Eq.~\eqref{ultralocal} in order to prevent non-local behaviour. At some scale $\Lambda$, dimension analysis gives
\begin{equation}
	G_{ij}^\Lambda
	=
	G_{IJ}(\varphi/\Lambda)
	\, \delta_\Lambda^{(4)}(x-x')
	\ ,
	\label{ultralocal2}
\end{equation}
where $\delta_\Lambda^{(4)}(x-x')$ is the regularized Dirac delta. A smooth Wilsonian cutoff $\Lambda$ can be implemented via the heat-kernel regularization~\footnote{Other cutoff implementations, such as a hard cutoff or lattice regularization with lattice spacing $\Lambda^{-1}$, give the same functional dependence on $\Lambda$. The numerical factor in the denominator $(2\pi)^{2}$ is however regularization-dependent.}:
\begin{equation}
	\delta_\Lambda^{(4)}(x)
	=
	\frac{\Lambda^4}{(2\pi)^{2}} e^{\frac{-x^2 \Lambda^2}{2}}
	\ ,
	\label{ultradiv}
\end{equation}
so that
\begin{equation}
    \delta_\Lambda^{(4)}(0) = \frac{\Lambda^4}{(2\pi)^{2}}
    \ .
    \label{delta}
\end{equation}

We shall be interested in the scalar sector $\varphi^i = (\varphi(x),\varphi^\dagger(x))$ with $i=(\varnothing, x)$. Restricting to quadratic inverse powers of the cutoff, we find
\begin{equation}
	\renewcommand{\arraystretch}{1.5}
	G_{IJ}
	=
	\begin{pmatrix}
		c_1 + c_2 \frac{\varphi^\dagger \varphi}{\Lambda^2} & c_3 + c_4 \frac{\varphi^\dagger \varphi}{\Lambda^2}
		\\
		c_5 + c_6 \frac{\varphi^\dagger \varphi}{\Lambda^2} & c_7 + c_8 \frac{\varphi^\dagger \varphi}{\Lambda^2}
	\end{pmatrix}
	+ \mathcal{O}(\Lambda^{-3})
	\ ,
	\label{metric}
\end{equation}
where $c_i$ are dimensionless free parameters. Note that there are no other terms invariant under $U(1)$ at this order. Eq.~\eqref{dimr} can be readily computed:
\begin{align}
    \Det G_{ij}
    &=
    \exp\left[\frac{\Lambda^4}{(2\pi)^{2}} \int\mathrm{d}^4x \, \log\left(A + B \frac{\varphi^\dagger \varphi}{\Lambda^2}\right)\right]
    \ ,
    \label{detcut}
\end{align}
where $A,B$ are merely redefinitions of $c_i$.
The integrand of \eqref{detcut} can be written as:
\begin{align}
	\Lambda^4 \log\left(A + B \frac{\varphi^\dagger \varphi}{\Lambda^2}\right)
	&=
	\Lambda^4 \log A
	+ \Lambda^2 \frac{B}{A} \varphi^\dagger \varphi
        \\
        &
	- \frac12 \frac{B^2}{A^2} (\varphi^\dagger \varphi)^2
	+ \mathcal{O}(\Lambda^{-2})
	\ .
	\label{Lexp}
\end{align}
We recall that in Wilson's effective field theory, the theory is finite by construction. However, even if one were to take $\Lambda\to\infty$ 
as in the ordinary view of renormalization, the positive powers of $\Lambda$ in Eq.~\eqref{Lexp} would get cancelled out by the counter terms, the negative powers of $\Lambda$ would go to zero, but the third term is cutoff-independent and remains finite after taking $\Lambda$ to infinity. Therefore,
\begin{equation}
	\Det G_{ij}
    =
    \exp\left[- \frac{1}{4 \pi^2} \frac{B^2}{A^2} \int\mathrm{d}^4x \, (\varphi^\dagger \varphi)^2 \right]
    \ ,
    \label{Fcorrection}
\end{equation}
which shows that the functional measure cannot be neutralized by dimensional regularization. The correction \eqref{Fcorrection} provides a finite contribution to the renormalization of the quartic interaction, thus changing its renormalization group running. The same would also happen for higher-order coupling constants should one include higher-order operators in the metric \eqref{metric}. In such an expansion, one thus recovers all possible interactions that are already present in standard effective field theory (without a functional measure), but their corresponding coefficients are shifted. Clearly, this is only possible at the analytical regime of the logarithm in Eq.~\eqref{detcut}. The effects of $A=0$, for example, cannot be reproduced by any finite order of the ordinary effective field theory expansion. They correspond to truly non-perturbative effects.

Plugging Eqs.~\eqref{detcut} and \eqref{measure} into~\eqref{Z}, one sees that the effect of this modified integration measure in field space can be rephrased as a shift in the effective potential, such that\footnote{Notice that eq.~\eqref{Z} involves the Euclidean action, so a positive contribution to $-S[\varphi]$, as in eq.~\eqref{detcut}, corresponds to a negative contribution to the effective potential.}~\cite{Kuntz:2024opj}
\begin{equation}
    V_\text{eff} = V_\text{cl} - \dfrac{\Lambda^4}{8\pi^2}\log\left(A+B\dfrac{\varphi^\dagger\varphi}{\Lambda^2}\right)
    \ ,
\end{equation}
where $V_\text{cl}$ is the classical scalar potential.
In this work we are interested in the cosmological consequences of this additional contribution to the scalar potential. Since we have added a minimal modification to the trivial metric, cf. eq.~\eqref{metric}, we will dub this the \emph{minimal modified functional measure model} (or MMFMM).

\section{Zero temperature effective potential and physicality conditions}
\label{sec:physicality}

Let us assume a Higgs-like field governed by a potential (at the classical level) given by
\begin{equation}
    V_\text{cl}(\varphi) = -\mu^2\varphi^\dagger \varphi + {\lambda} (\varphi^\dagger \varphi)^2.
\end{equation}
Quantum corrections to this potential include the typical Coleman-Weinberg terms at 1-loop level, and also the quantum correction due to the modified integration measure. Apart from an overall constant, the latter can be rewritten such that the 1-loop zero-temperature effective potential becomes
\begin{equation}\begin{split}
  V^{(1)}(\phi, T=0) & = -\frac{\mu ^2}{2}\phi ^2 + \frac{\lambda}{4}\phi^4 \\
    & + \sum_i (-1)^{s_i} n_i\frac{m_i(\phi)^4}{64\pi^2}\bigg(\ln\frac{m_i^2(\phi)}{Q^2} - \frac{1}{2}\bigg)\\
    & - \frac{\Lambda^4}{8\pi^2}\ln\bigg(1 - C\frac{\phi^2}{\Lambda^2}\bigg)
\label{eq:V1}
\end{split}\end{equation}
where the sum in $i$ runs over all vector bosonic ($s_i=0$) and fermionic ($s_i=1$) degrees of freedom running in the loops, and we defined $\varphi \equiv \phi/\sqrt{2}$ and $C\equiv -B/2A$. For the mass of the $W$ boson we take $80.36~\text{GeV}$, while for the $Z$ boson we use $91.19~\text{GeV}$, and for the top quark $173.1~\text{GeV}$. The contributions from the masses of other particles are negligible and were therefore disregarded.

{The parameters $\mu$ and $\lambda$ are fixed at the electroweak scale} by imposing that the Higgs potential has a minimum at $v = 246~\text{GeV}$ and that its mass is $m_h = 125~\text{GeV}$, i.e. $\partial V^{(1)}/\partial\phi|_{\phi=v} = 0$ and $\partial^2 V^{(1)}/\partial\phi^2|_{\phi=v} = m_h^2$~\footnote{ The second derivative of the effective potential coincides with the physical mass, which is defined by the propagator's pole. By means of Eq.~\eqref{RG}, the physical mass $m_h$ is cutoff-independent.}.
Thus the full effective potential can be rewritten as
\begin{equation}\begin{split}
  V^{(1)}(\phi,0) & = \dfrac{m_h^2}{8v^2}(\phi^2-v^2)^2 \\
    & + \sum_i \dfrac{n_ig_i^4}{64\pi^2}\left(\phi^4\left(\ln\dfrac{\phi^2}{v^2}-\dfrac{3}{2}\right) +2\phi^2v^2\right)   \\
    & - \frac{\Lambda^4}{8\pi^2}\ln\bigg(1 - C\frac{\phi^2}{\Lambda^2}\bigg)  \\
    &- \dfrac{C\Lambda^2}{4\pi^2}\dfrac{1-2Cv^2/\Lambda^2}{(1-Cv^2/\Lambda^2)^2}\dfrac{\phi^2}{2}\\
    &- \dfrac{C^2}{4\pi^2(1-Cv^2/\Lambda^2)^2}\dfrac{\phi^4}{4}.
\label{eq:Veff}
\end{split}\end{equation}

\begin{figure}
    \centering
    \includegraphics[width=\linewidth]{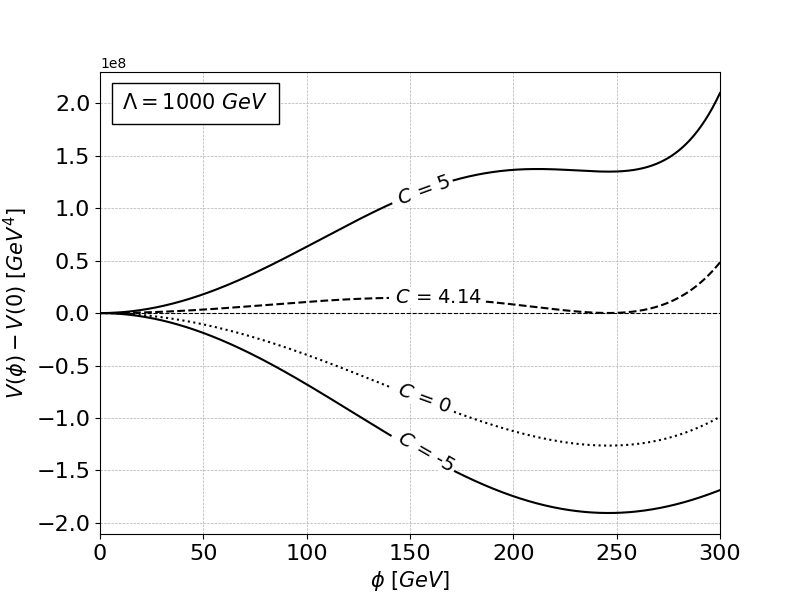}
    \caption{Normalized shapes of the potential at zero temperature for different values of $C$ for a given $\Lambda = 1000$~GeV. The dotted line ($C = 0$) corresponds to the SM potential. For $C < 0$ the curve is always below the SM one. For $C>0$ the curve rises above the SM in the range $\phi < \Lambda/\sqrt{C}$. Note that for $C>C_c(\Lambda = 1000~\text{GeV})\approx 4.14$ the symmetry-breaking vacuum is not a global minimum and spontaneous symmetry breaking would not take place.}
    \label{fig:Veff_Cs}
\end{figure}

The shape of this modified potential is determined by the free parameters $C$ and $\Lambda$. The latter is a scale that sets the energy at which the effect of the modified functional measure starts to become sensitive. On the other hand, $C$ parameterizes the deviation from the usual case, when the measure is trivial. Indeed, it is easy to notice that the Standard Model Higgs potential is recovered for $C = 0$. The role of $C$ for a fixed value of $\Lambda$ is exemplified in figure~\ref{fig:Veff_Cs}.

To better understand the impact of these new contributions to the effective potential, let us denote by $V_\text{measure}(\phi)$ the new terms arising from the modified measure, i.e. the last three rows of eq.~\eqref{eq:Veff}. Denoting $t\equiv C\phi^2/\Lambda^2$ and $t_0\equiv C v^2/\Lambda^2$ to simplify the notation, it is easy to see that
\begin{equation}
    V_\text{measure} = -\dfrac{\Lambda^4}{8\pi^2}\left[ 
        \ln(1-t) + \dfrac{t^2 + 2t - 4tt_0}{2(1-t_0)^2}
    \right],
    \label{eq:Vmeasure}
\end{equation}
and that the derivative of this potential has the form
\begin{equation}
    \dfrac{\partial V_\text{measure}}{\partial\phi} = \dfrac{\Lambda^2}{4\pi^2} C\phi \dfrac{(t-t_0)^2}{(1-t)(1-t_0)^2}.
\end{equation}
For $C>0$ and $0<\phi<\Lambda/\sqrt{C}$, one has $0<t<1$ and $\partial V_\text{measure}/\partial \phi>0$. Since $V_\text{measure}(0)=0$, this means that $V_\text{measure}$ will always lead to a \emph{positive} contribution to the effective potential in this range of $\phi$, thus \emph{raising} it relative to the SM case. Moreover, this $\phi$ range must include the minimum $\phi=v$ in order for the logarithm to be real up to this point, as we will discuss shortly. Thus, the minimum of the potential is also raised as $C$ increases. It is well-known that this tends to lead to stronger first-order phase transitions in the early Universe~\cite{Dorsch:2017nza}. For this reason we will henceforth limit ourselves to the case\footnote{We have explicitly checked that no strongly first-order phase transitions occur for $C<0$.} $C>0$.

One can immediately impose a few constraints on these parameters from simple consistency conditions of the theory and also from LHC bounds on the scalar sector. 

\textbf{Existence of electroweak symmetry breaking:} 
From figure~\ref{fig:Veff_Cs} one sees that, for a fixed $\Lambda$, there is  a certain critical value $C_\text{noEWSB}(\Lambda)$ above which the global minimum is at $\phi=0$ rather than at $\phi=v$, even at zero temperature. This would mean that symmetry breaking would never have taken place. This should be avoided, so we impose 
\begin{equation}
    C<C_\text{noEWSB}(\Lambda).
    \label{eq:Cc}
\end{equation}

\textbf{Stability of the field configuration with vev $\phi$}: By simple inspection one can see that, for $C>0$, the zero-temperature potential becomes complex-valued for large enough values of the field $\phi$. This means that the quantum state whose vev is this $\phi$ will actually be an unstable state and will eventually decay~\cite{Weinberg:1987vp}. In order to avoid such complications in the field range of our interest, we require the argument of the logarithm to be positive at least for $\phi\leq \phi_\text{stab}$ with $\phi_\text{stab}\geq v$. Fixing this $\phi_\text{stab}$ then gives an upper bound
\begin{equation}
    C<\left(\dfrac{\Lambda}{\phi_\text{stab}}\right)^2.
\end{equation}

\textbf{Vacuum stability:} In principle one could also wonder about the possibility that the broken vacuum $\phi=v$ is not the deepest minimum of the potential, which would lead to the possibly catastrophic situation of an unstable vacuum. From eq.~\eqref{eq:Veff} one sees that, for $C>0$, this may indeed be the case (though not necessarily), since for large field values the potential becomes effectively
\begin{equation}
    V^{(1)}(\phi\to\infty,0) \to \lambda_\text{eff}\dfrac{\phi^4}{4},
\end{equation}
with
\begin{equation}
    \lambda_\text{eff} = \dfrac{m_h^2}{2v^2} +\sum_i \dfrac{n_i g_i^4}{16\pi^2} \ln\dfrac{\phi^2}{v^2} 
    - \dfrac{C^2}{4\pi^2}\left(1-\dfrac{Cv^2}{\Lambda^2}\right)^{-2}.
    \label{eq:leff}
\end{equation}
However, we have shown above that, in the range $0<\phi < \Lambda/\sqrt{C}$, the new terms $V_\text{measure}$ can only \emph{raise} the SM potential. So the unboundedness of $V^{(1)}$ from below will only be relevant in the region where $V^{(1)}$ becomes complex and the state with vev $\phi$ is itself unstable, as discussed above. Moreover, the real part of the potential, which describes the energy density of the field configuration, displays an arbitrarily large barrier before we enter the problematic unstable region, so tunneling to the unstable state is highly suppressed. For this reason we will not pursue the consequences of this condition any further.

\textbf{Bounds on the Higgs self-coupling:}
The model in consideration adds corrections to the Higgs trilinear coupling. The Higgs $H = \phi - v$ has a cubic self-interaction through the term $\lambda v H^3$, where, in the Standard Model, $\lambda_{\rm SM} = m_h^2 / (2v^2)$. Our model modifies this value, such that an effective trilinear coupling is achieved with $\lambda_{eff} = \kappa_\lambda\lambda_{\rm SM}$. Using the fact that 
\begin{equation}
    \lambda_{eff} = \dfrac{1}{6v}\left.\dfrac{\partial^3 V}{\partial\phi^3}\right|_{\phi=v}
\end{equation}
we obtain
\begin{equation}
    \kappa_\lambda = 1+
    \sum_i\left(\dfrac{n_ig_i^4}{24\pi^2}\dfrac{v^2}{m_h^2}\right) + 
    \dfrac{2 v^4}{3\pi^2m_h^2 \Lambda^2}
        \left(  
        \dfrac{C}{1 - \frac{Cv^2}{\Lambda^2}}
        \right)^3.
    \label{eq:kappa}
\end{equation}
The current 68\% C.L. limits on the Higgs trilinear coupling are $1.1\leq \kappa_\lambda \leq 4.8$~\cite{ATLAS:2022jtk}. These were obtained from dihiggs + single higgs production while keeping all the other couplings fixed at the SM values, as is appropriate for the case under consideration here.

\begin{figure}
    \centering
    \includegraphics[width=\linewidth]{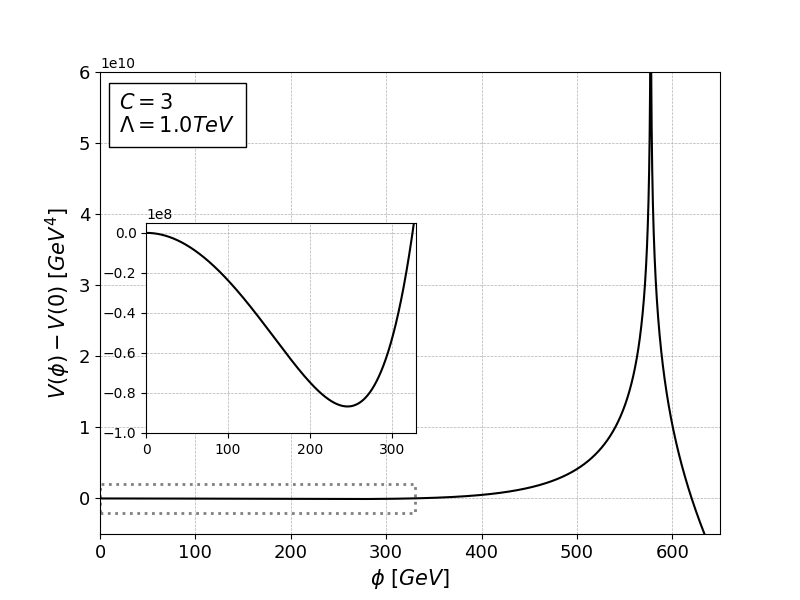}
    \caption{Normalized shape of the potential at zero temperature for $\Lambda = 1~\text{TeV}$ and $C = 3$. The region in the dotted line correspond to the area in the zoom. In the figure one can see the infinity barrier at $\phi = \frac{\Lambda}{\sqrt{C}}\approx 577~\text{GeV}$ that assures that a transition will never occurs after achieved the currently vacuum value. }
    \label{fig:Vedd Lambs}
\end{figure}

\section{Thermal effective potential and the electroweak phase transition}
\label{sec:Vth}

We now consider the effects of finite temperature on the effective potential. The 1-loop thermal correction is well known in the literature~\cite{Laine:2016hma, Hindmarsh:2020hop, Espinosa:1992kf} and is given by
\begin{equation}\begin{split}
    \Delta V^{(1)}(\phi, T) & = \dfrac{T^4}{2\pi^2} \displaystyle \sum_{i=\text{boson}}n_iJ_B\left(\dfrac{m_i(\phi)^2}{T^2}\right) \\ 
    & - \dfrac{T^4}{2\pi^2} \sum_{i=\text{fermion}}n_iJ_F\left(\dfrac{m_i(\phi)^2}{T^2}\right) \\
    & - \dfrac{T}{12\pi}\cdot 3\left({m}_L^3(\phi,T) - m_L^3(\phi,0)\right),
    \label{eq:Vth}
\end{split}\end{equation}
where we are again summing over all fermionic and bosonic contributions, and $J_{B,F}$ are numerically evaluated functions defined by the integrals
\begin{equation}
    J_{B,F}(y) = \int_0^\infty x^2 \ln(1\mp e^{-\beta\sqrt{x^2 + y^2}}),
\end{equation}
with $-$ (resp. $+$) corresponding to bosonic (resp. fermionic) contributions. The third line in eq.~\eqref{eq:Vth} are the so-called daisy terms, which introduce thermal corrections to the mass of the longitudinal components of the gauge bosons in order to improve the loop expansion. For our model, with a SM-like particle content, we have \cite{Espinosa:1992kf} ${m}_L^2(\phi,T) = \frac{1}{4}\overline{g}^2\phi^2 + \frac{11}{6}\overline{g}^2T^2$ with $\overline{g}^2 = 4(m_W^2(v) + m_Z^2(v))/(3v^2)$.

The full effective potential is thus:
\begin{equation}
    V^{(1)}(\phi,T) = V^{(1)}(\phi,T=0) + \Delta V^{(1)}(\phi,T).
\end{equation}

In the early universe, for sufficiently high temperatures, the thermal part of the potential is proportional to $+T^2\phi^2$ and dominates over the negative quadratic terms in eq.~\eqref{eq:Veff}. In this case all contributions to the effective potential are positive, and there is only one minimum at $\phi=0$: one says that electroweak symmetry is restored. As the Universe cools down, eventually a second minimum appears and becomes energetically degenerate with $\phi=0$: this is the so-called critical temperature of the electroweak phase transition, $T_c \sim \mathcal{O}(100)$~GeV. Below this temperature the Universe goes through a phase transition in which the Higgs vev $\phi$ acts as the order parameter. It is known that, for the Standard Model, this process is actually a smooth crossover~\cite{DOnofrio:2015gop}. We will show that the MMFMM could lead to a strong first order phase transition for sufficiently large $C>0$.

With that in mind, we proceed to compute the relevant parameters following the standard procedures~\cite{Hindmarsh:2020hop, Caprini:2015zlo, Caprini:2019egz}. Below the critical temperature there will be a probability that the Higgs field will transition from the symmetric state $\phi_s = 0$ to the broken state $\phi_b \neq 0$. This will happen via the nucleation of bubbles of the broken phase in a universe filled with the symmetric state. The rate at which the bubbles will be nucleated per unit volume is
\begin{equation}
    \Gamma \approx A(t)e^{-S_3/T},
    \label{eq:rate}
\end{equation}
with the 3D Euclidean action
\begin{equation}
    S_3[\phi] = \displaystyle\int {\rm d}^3x \left[\dfrac{1}{2}(\nabla\phi)^2 + \left(V(\phi, T) - V(0,T)\right)\right].
    \label{eq:S3}
\end{equation}
In principle the rate in eq.~\eqref{eq:rate} would include a sum over all configurations mediating the transition between the two vacua,  but because of the exponential suppression the dominant contribution will come from the configuration that actually minimizes the action~\cite{Mukhanov:2005sc}. It can also be shown that the minimal energy configuration is spherically symmetric~\cite{Coleman:1977fateOfTheFalseVacuum}, so the Euler-Lagrange equations for the configuration that minimizes the action~\eqref{eq:S3} is 
\begin{equation}
    \dfrac{\rm d^2 \phi}{\rm dr^2} + \dfrac{2}{r}\dfrac{\rm d\phi}{\rm dr} - \dfrac{\partial V}{\partial \phi}(\phi,T) = 0 
    \label{eq:bounce}
\end{equation}
with boundary conditions
\begin{equation}
    \phi'(0) = 0 \;,\; \phi(0) \neq 0 \;,\; \phi(r\to\infty) = 0
\end{equation}
to ensure that the configuration indeed interpolates the two vacuum states.

We solve this equation using a custom-made Python code implementing a shooting method, with a bisection algorithm applied for finding the initial condition $\phi(0)$ which will lead to the desired behaviour of $\phi(r\to \infty)= 0$. Having found a solution, one can plug it back into Eqs.~\eqref{eq:S3} and~\eqref{eq:rate} to find the nucleation rate per unit volume, which then allows us to compute the number of nucleated bubbles per Hubble horizon. 
The nucleation temperature $T_n$ is defined as the temperature at which this number is unity. 
For phase transitions occurring at the electroweak scale $\mathcal{O}(100~\text{GeV})$ this condition essentially reduces to~\cite{Caprini:2019egz}
\begin{equation}
    \dfrac{S_3(T_n)}{T_n} \approx 140.
    \label{eq:ST_140}
\end{equation}
This is the criterium we use to find $T_n$. 

There are two other parameters that can be calculated from equilibrium considerations alone (i.e. from the knowledge of the effective potential and the relevant transition temperature). One is often called the transition strength parameter, and is a measure of the energy budget of the transition:
\begin{equation}
    \alpha = \dfrac{1}{\rho_r}\Delta\left.\left(V^{(1)}(\phi,T) - \dfrac{1}{4}T\dfrac{\partial V^{(1)}}{\partial T}  \right)\right|_{T=T_n}
    \label{eq:alpha}
\end{equation}
where $\Delta$ corresponds to the difference between the expression in parenthesis at the symmetric and the broken minima at $T_n$, and $\rho_r = g_{*}\, \pi^2 T_n^4/30\,$ is the energy density due to radiation from relativistic species\footnote{Since we are not altering the Standard Model field content, the number of relativistic degrees of freedom at the electroweak phase transition is $g_* = 106.75$.}. Notice that the quantity in parenthesis is the trace anomaly $\theta = $ (trace of energy-momentum tensor of the plasma)$/4 = (e - 3p)/4$, where $e=T\partial p/\partial T - p$ is the energy of the plasma and $p=-V^{(1)}(\phi,T)$ its pressure. The definition of this parameter varies across the literature, with the trace anomaly being replaced by either $e$ or $-p$ in some cases\footnote{Refs.~\cite{Athron:2024_cosmologicalPhaseTransitions,Giese:2020rtr} also discuss the possibility of using another generalization of $\alpha$, by letting the trace anomaly depend on the speed of sound: $\tilde{\theta} = 1/4(e-p/c_s^2)$. This reduces to our definition when $c_s = 1/\sqrt{3}$, the speed of sound obtained from the bag model for the plasma, which is the case considered here.}~\cite{Giese:2020rtr} and the denominator $\rho_r$ being replaced by $3/4$ of the enthalpy $3w_f/4$ in other cases.  A discussion of these different uses can be found in \cite{Athron:2024_cosmologicalPhaseTransitions}. 

Another relevant parameter is the (inverse) characteristic time scale of the transition, which can be computed as~\cite{Caprini:2015zlo, Schmitz:2020syl}
\begin{equation}
    \dfrac{\beta}{H_*} = T_n \left.\dfrac{dS_3}{dT}\right|_{T=T_n}.
    \label{eq:beta}
\end{equation}
From this one computes the average size of the colliding bubbles as $R_*\equiv (8\pi)^{1/3} v_w \beta^{-1}$ (with $v_w$ the bubble expansion velocity)~\cite{Enqvist:1991xw}, so one expects the GW spectrum to be proportional to $v_w\beta^{-1}$.

The calculation of the bubble expansion velocity depends on an adequate estimate of the counter-pressure against the expanding wall, which has an equilibrium as well as a non-equilibrium contribution. To find the latter one needs to solve the integro-differential Boltzmann equation, which is only well-defined once we know how to compute the collision terms. There are various approaches to this issue, ranging from a fully numerical solution of the Boltzmann system on a lattice~\cite{DeCurtis:2022hlx}, an expansion of the non-equilibrium fluctuations in Chebyshev polynomials~\cite{Laurent:2022jrs} allied with a collocation method for solving the Boltzmann system (and a fully numerical computation of the collision terms), or a fluid Ansatz for the fluctuations~\cite{Moore:1995si, Dorsch:2023tss, Dorsch:2024jjl} (which allows for an analytic calculation of the collision terms at leading-log~\cite{Dorsch:2021ubz}) together with a procedure for taking moments to reduce the Boltzmann equation to a linear system of ODEs. Regardless of the approach chosen, the full setup also requires solving the hydrodynamical equations to determine whether the wall front expands as a deflagration, a hybrid or a detonation~\cite{Espinosa:2010hh}. Overall, regardless of which approach is chosen, one can safely say that finding $v_w$ is a highly non-trivial matter\footnote{See ref.~\cite{Ekstedt:2024fyq} for a recent numerical code for computing the collision terms and finding $v_w$ for any given model, using the collocation method and an expansion in Chebyshev polynomials.}. In this work we assume for simplicity that $v_w\approx 1$, which is typically within the correct order of magnitude of full results for models with a Standard Model particle content and transitions strengths of the order considered here~\cite{Dorsch:2023tss, Dorsch:2024jjl}.

\section{Gravitational Waves}
\label{sec:GW}

After nucleating, bubbles expand and eventually collide, filling the entire Universe with the broken phase. This happens at a characteristic temperature called percolation temperature $T_* < T_n$, defined as the temperature at which about $30\%$ of the Universe is filled by the true vacuum~\cite{Athron:2022mmm}. These collisions break the sphericity of the bubbles, inducing a time-varying quadrupole moment of energy-momentum which sources gravitational waves. It is at this point that a stochastic GW spectrum is produced. This means that the relevant temperature for computing the phase transition parameters should actually be the percolation temperature $T_*$. However, one typically has $T_* \approx T_n$, and this is the approximation we will use, which holds for non-supercooled transitions like the ones considered here. For this reason Eqs.~\eqref{eq:alpha} and~\eqref{eq:beta} have been defined at $T_n$.

For phase transitions that nucleate and percolate at a finite temperature, there are two main sources of gravitational waves\footnote{In principle there would also be a contribution from the kinetic energy in the scalar field itself. This is most important in the case of runaway bubbles, when the counter-pressure is never enough to stop the walls from accelerating indefinitely. Recent calculations show that this happens less often than previously expected~\cite{Hoche:2020ysm, Azatov:2020ufh, Ai:2024shx, Long:2024sqg}, so we choose to neglect the scalar contribution to the spectrum.}, namely sound waves and turbulence (i.e. linear and nonlinear perturbations) in the plasma~\cite{Caprini:2015zlo, Caprini:2019egz, Schmitz:2020syl}. It is known that these spectra satisfy a broken power-law and have a peak at a characteristic frequency $\sim\mathcal{O}(\text{few mHz})$, so we will parametrize it as
\begin{equation}
    h^2\Omega(f) = h^2 \Omega^\text{(peak)}\times S(f),
\end{equation}
with $\Omega$ the ratio of energy density in GWs over the critical density of our Universe. This critical density is proportional to the square of the Hubble constant today, $H_0^2 = (100 h~\text{km/s/Mpc})^2$. The value of $h$ still carries a large uncertainty due to tensions in measurements from cosmological data and from supernovae. To avoid the impact of these uncertainties on the GW spectrum, it is common to actually write the GW energy density in terms of $h^2 \Omega$ rather than $\Omega$ itself.

For the sound wave contribution the spectrum has been determined from numerical simulations\footnote{These simulations have only taken into account sufficiently weak phase transitions, with $\alpha\ll 1$. Whether and how the spectra would change for stronger transitions remains an open issue. However, as will be seen in figure~\ref{fig:ab} below, in this model we are most often in the region of $\alpha\lesssim 1$.}~\cite{Hindmarsh:2017gnf} and encapsulated in the fit~\cite{Caprini:2015zlo}
\begin{equation}\begin{split}
    h^2\Omega_{\rm sw}^\text{peak}(f)&=2.65\times10^{-6}\,\times\\
    &\times\left(\dfrac{v_w}{\beta/H_*}\right)\left(\frac{\kappa_\text{sw}\alpha}{1+\alpha}\right)^2 \left(\frac{100}{g_*}\right)^\frac{1}{3}\mathcal{Y}_\text{sup}
    \label{eq:sw spectrum}
\end{split}
\end{equation}
for the peak amplitude. Apart from the parameters already described above, the peak amplitude also depends on the efficiency in converting the energy released by the transition into sound waves, $\kappa_\text{sw}$. We will use the one for non-runaway phase transitions \cite{Schmitz:2020syl, Caprini:2015zlo}: 
\begin{equation}
    \kappa_{\rm sw} = \dfrac{\alpha}{0.73 + 0.83\sqrt{\alpha} + \alpha}.
\end{equation}
Moreover, in equation~\eqref{eq:sw spectrum} we have also included a suppression factor $\mathcal{Y}_\text{sup}$ (as compared to the result quoted in ref.~\cite{Caprini:2015zlo}) to account for the possibility that the sound waves die off into turbulence before a Hubble time. This was not included in the initial simulations of refs.~\cite{Hindmarsh:2013xza, Hindmarsh:2017gnf}, which assume long-lasting sound waves, but it turns out that in typical models this is not a reasonable assumption and sound waves will typically decay at a characteristic time~\cite{Ellis:2018mja, Ellis:2020awk}
\begin{equation}
    \tau_\text{sw}H_* = (8\pi)^{\frac{1}{3}}\dfrac{v_w}{\beta/H_*}\sqrt{\dfrac{4}{3}\dfrac{1+\alpha}{\kappa_{\rm sw}\alpha}}.
\end{equation} 
We will then model this suppression factor as
\begin{equation}
    \mathcal{Y}_\text{sup} = \min\{1, \tau_{\rm sw}H_*\}.
\end{equation}

Equation~\eqref{eq:sw spectrum} provides only an expression for the peak of the spectrum. Its shape is given by
\begin{equation}
    S_{\rm sw}(f)=\left(\dfrac{f}{f _{\rm sw}}\right)^{3}\left(\frac{7}{4+3(f/f _{\rm sw})^{2}}\right)^{7/2},
    \label{eq:Ssw}
\end{equation}
with $f_{\rm sw}$ the redshifted frequency of the peak frequency as measured today,
\begin{equation}
\begin{aligned}
     f_{\rm sw} &= 1.9e-2\,\text{mHz}\,\times\\
     &\quad\times \frac{1}{v_w}\left(\dfrac{\beta}{H_*}\right)\left(\frac{T_*}{{100}\,\text{GeV}}\right)\left(\frac{g_*}{100}\right)^\frac{1}{6}.
\end{aligned}
\end{equation}

The GW spectrum produced from turbulence in the plasma is often computed in the literature using the fit found in ref.~\cite{Caprini:2015zlo},
\begin{align}
\begin{aligned}
    h^2\Omega_\text{turb}^\text{peak}(f) &= 3.35\times 10^{-4}\,\times \\
    &\times \left(\dfrac{v_w}{\beta/H_*}\right)\left(\frac{\kappa_{\rm turb}\alpha}{1+\alpha}\right)^\frac{3}{2} \left(\frac{100}{g_*}\right)^\frac{1}{3},
\end{aligned}
\end{align}
with shape
\begin{equation}
    S_{\rm turb}=\frac{(f/f _{\rm turb})^{3}}{[1+(f/f_{\rm turb})]^{\frac{11}{3}}\left(1+8\pi\frac{f}{h_*}\right)}. 
    \label{eq:Sturb}
\end{equation}
The $h_*$ term in the equation corresponds to the Hubble parameter at the epoch of the phase transition, rescaled by the cosmological redshift to relate it to the present time, and is defined as
\begin{equation}
    h_*=16.5\times10^{-3}mHz\left(\dfrac{T_*}{100Gev}\right)\left(\frac{g_*}{100}\right)^{\frac{1}{6}},
\end{equation}
and $f_\text{turb}$ is the peak frequency measured today,
\begin{equation}
\begin{aligned}
     f_{\rm turb} & = {2.7e-2}\,\text{mHz}\,\times  \\
     &\times \frac{1}{v_w}\left(\dfrac{\beta}{H_*}\right)\left(\frac{T_*}{{100}\,\text{GeV}}\right)\left(\frac{g_*}{100}\right)^\frac{1}{6}.
\end{aligned}
\end{equation}

We emphasize, however, that the GW spectrum from turbulence is actually still a matter of intense investigation~\cite{Caprini:2019egz}. Even if one is willing to use the expression above, one would need to know  the efficiency factor $\kappa_\text{turb}$, corresponding to the fraction of released energy that is actually converted to turbulence.
There is so far no appropriate modeling for this parameter. It is sometimes assumed that $\kappa_{\rm turb} =\varepsilon\kappa_{\rm sw}$, with typical values ranging from $5 - 10 \%$~\cite{Caprini:2015zlo, Schmitz:2020syl, Alves:2019igs}. Sometimes extreme cases as $\varepsilon=0$ (negligible turbulence)~\cite{Caprini:2019egz} and $\varepsilon=1$\cite{Ellis:2019oqb} are also assumed. In the next section we will show how the spectrum changes if we vary $\varepsilon$ from $0$ to $1$. 

\subsection*{Analysis of detectability}

To assess the detectability of these GWs one must calculate the \textit{signal-to-noise ratio} (SNR), obtained by comparing the predicted signal, $\Omega_\text{theory}(f)$, to the noise spectrum of a given interferometer, $\Omega_\text{noise}(f)$, and integrating this ratio over the operational time of the experiment. Moreover, due to the broadband nature of the stochastic GW spectrum from phase transitions, it is also important to integrate this ratio over all frequencies accessible at the experiment, since an accumulation of small $\Omega_\text{theory}(f)/\Omega_\text{noise}(f)$ could still yield a large enough SNR when summed over. If the spectrum has a power-law shape, $\Omega_\text{theory}(f) = \Omega^\text{peak} f^n$, ref.~\cite{Thrane:2013oya} provided a recipe for constructing the so-called \emph{power-law integrated sensitivity curves} (PLISCs) by varying the power $n$ and calculating the value of $\Omega^\text{peak}$ for which the spectrum would be detectable (SNR $>1$). This would yield an envelope of straight lines in a log-log plot of $\Omega$ vs $f$, such that the actual spectrum (if it has a power-law shape) would be detectable if its curve lies \emph{above} this envelope. This provides a convenient graphical way of displaying detectability results.

But from eqs.~\eqref{eq:Ssw} and~\eqref{eq:Sturb} one sees that the spectrum from cosmological phase transitions is not a simple power-law, but a broken power law with a peak. In this case the comparison of the spectrum with the PLISCs would only be plausible if the peak lies beyond the range of detectability of the experiment, such that the spectrum would effectively have a simple power-law shape as far as this experiment is concerned. This is a too restrictive requirement, and it would be useful to have another reliable, simple and graphical way to assess the detectability of a generic broken power-law spectrum as predicted by phase transitions. 

Now, because the shape of these spectra is known, given e.g. by eqs.~\eqref{eq:Ssw} and~\eqref{eq:Sturb}, the integral of $\Omega_\text{theory}(f)/\Omega_\text{noise}(f)$ can be readily performed, and one can rewrite the condition SNR~$>1$ in terms of a condition on the minimal $\Omega^\text{peak}$ that would yield the spectrum detectable (as a function of the peak frequency). This new method was recently developed in ref.~\cite{Schmitz:2020syl} and is dubbed the \emph{peak integrated sensitivity curves} (PISCs). This provides a reliable graphical way of displaying detectability results for cosmological GW backgrounds. The disadvantage, compared to the PISCs, is that this method is only applicable for a comparison of the peak amplitude, and all information on the spectral shape has been integrated over.

In what follows we will use mostly the PISC method for checking an SNR~$>1$, except in one case when the full spectral shape is shown for illustrative purposes. We will then also use this opportunity to briefly emphasize the advantage of the PISC over the PLISC method for studying cosmological phase transitions.

\section{Results}
\label{sec:results}

Our task is to evaluate the possibility of using gravitational wave detectors to probe deviations in the scalar potential arising from terms of the form $\sim \Lambda^4\log(1-C\phi^2/\Lambda^2)$ (cf. eq.~\eqref{eq:Vmeasure}). For this purpose, we have performed a scan over the $(\Lambda, C)$ parameters, with $\Lambda\in (500, 3000)$~GeV, evaluating the phase transition parameters and the GW spectrum. 

\begin{figure}
    \centering
    \includegraphics[width=0.9\linewidth]{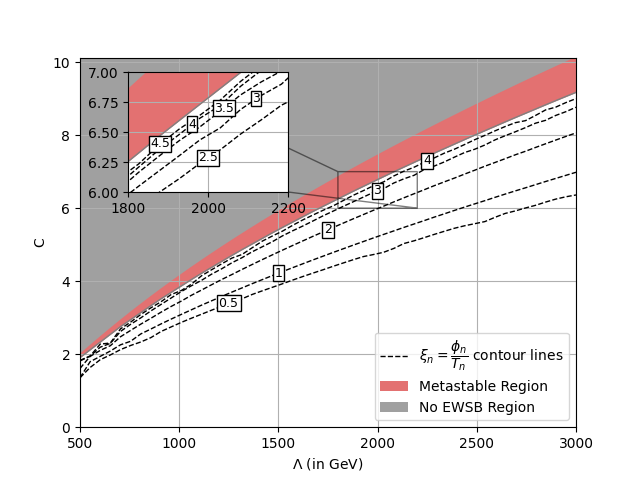}
    \caption{Contour plot for the values of the transition strength parameter $\xi_* = \phi_*/T_*$ in the $(\Lambda, C)$ plane. The figure illustrates clearly how increasing $C$ leads to stronger phase transitions, as expected from the raising of the zero-temperature minimum shown in figure~\ref{fig:Veff_Cs}.
    }
    \label{fig:phi_T}
\end{figure}

Figure~\ref{fig:phi_T} illustrates the behaviour of the ratio $\phi_n/T_n$, i.e. the order parameter (Higgs VEV) per temperature at the start of bubble nucleation, which serves as a measure of how strongly first order the phase transition is. As expected, for fixed $\Lambda$ the transition becomes stronger as we increase $C$. This is because for $C>0$ the minimum of the potential is raised compared to the SM case, as discussed previously in section~\ref{sec:physicality}, so we need smaller temperature corrections to reach the critical (and also the nucleation) temperature. Since the VEV also decreases with $T$, a smaller transition temperature means larger $\phi/T$~\cite{Dorsch:2017nza}.

If $C$ is large enough one finds that the condition $S_3/T\approx 140$ in eq.~\eqref{eq:ST_140} is never satisfied. This means we never guarantee the nucleation of one bubble per Hubble horizon. We interpret this as a condition that the transition will not complete, i.e. the false vacuum is \emph{metastable}\footnote{Note that, in a radiation-dominated Universe, bubbles might expand faster than the Hubble rate, so a bubble might grow beyond the Hubble volume at which it was nucleated, and the transition could complete even with less than one bubble per horizon~\cite{Athron:2022mmm}. The condition of metastability is actually less stringent, namely that there must be less than $\sim \mathcal{O}(0.1)$ bubbles per horizon~\cite{Athron:2022mmm}, which in practice changes very little the exclusion regions we obtain here. Thus we choose to keep the usual condition of at least one nucleated bubble per horizon.}. As we increase $C$ even further we eventually raise the electroweak minimum above the symmetric phase even at $T=0$: this means that symmetry breaking would never have taken place. The \emph{metastability} region is shown in red in Fig.~\ref{fig:phi_T} (and in light gray in the following figures), and the region of \emph{no electroweak symmetry breaking} is shown in dark gray here and in the following plots.

\begin{figure}
    \centering
    \includegraphics[width=0.83\linewidth]{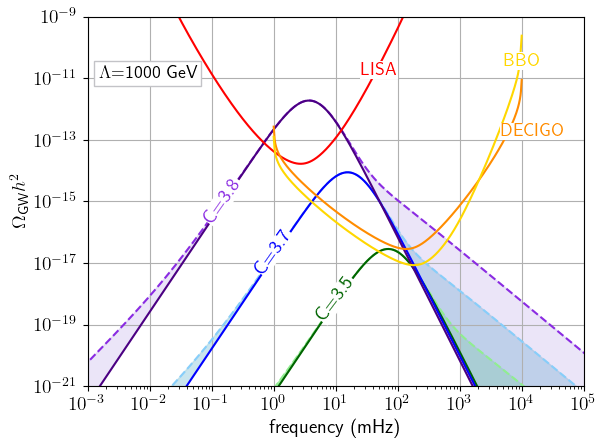}
    \caption{Typical gravitational wave spectra for a fixed value of $\Lambda=1000$~GeV and various $C$. The spectra exhibit 
    contributions from two different sources: the solid lines represent contributions from sound waves, while the dashed lines illustrate the contribution from turbulence assuming an efficiency factor $\varepsilon = 1$ (see text). The shaded region is the range obtained by varying $\varepsilon$ from $0$ to $1$. Sensitivity curves of the different GW detectors are power-law-integrated sensitivity curves (PLISCs)~\cite{Thrane:2013oya} taken from \cite{Schmitz:2020syl}. See the main body of the text for a discussion on these sensitivity curves.}
    \label{fig:sw+t}
\end{figure}

Once a certain point $(\Lambda,C)$ passes all the physicality constraints discussed in section~\ref{sec:physicality}, we can calculate the transition parameters and the GW spectrum. Figure~\ref{fig:sw+t} illustrates typical shapes of this spectrum for a fixed $\Lambda=1000$~GeV and differing values of $C$. As expected, the amplitude increases with $C$, because the transition gets stronger. Solid lines correspond to the contribution of sound waves only, whereas dashed lines include the contribution from turbulence with an extreme efficiency factor of $\kappa_\text{turb}=\kappa_\text{sw}$ (or $\varepsilon=1$). The result shows that turbulence changes the spectrum mostly in its tails, but does not significantly affect the peak. For this reason, and because any choice of $\varepsilon$ would be rather arbitrary, in the following discussions we will instead completely neglect the turbulence contribution, setting $\varepsilon=0$. Note that this will lead to a \emph{more conservative} estimate of how sensitive GW detectors are to this model.

We also show in figure~\ref{fig:sw+t}  the power-law integrated sensitivity curves (PLISC)~\cite{Thrane:2013oya} of three different detectors, namely LISA, DECIGO and BBO. As illustrated, it is not typically true that the spectrum has a simple power law shape for the whole detectability range of
these experiments. Thus, looking at figure~\ref{fig:sw+t} alone is in principle not enough to judge whether the spectrum is detectable\footnote{We reinforce that these PLISCs were constructed by assuming that the spectrum would keep increasing for all $f$. The fact that the power law is broken, and that the spectrum at some point decreases, means that the integral of $\Omega_\text{theory}(f)/\Omega_\text{noise}(f)$ would actually be smaller than for a power-law, and the comparison with the PLISC is not totally appropriate.}. We will henceforth use the PISC method to assess the detectability of these GWs.

\begin{figure}
    \centering
    \includegraphics[width=0.9\linewidth]{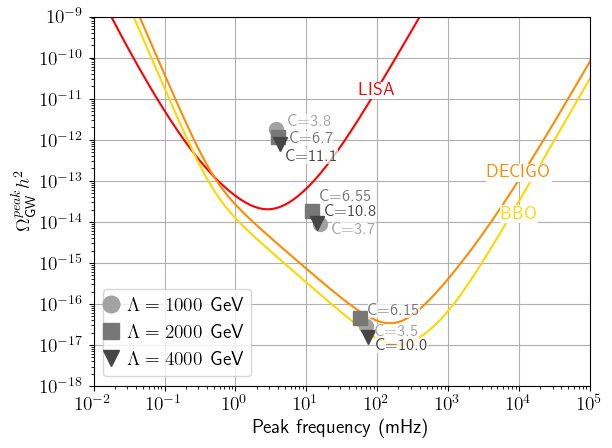}
    \caption{Peak amplitudes of the gravitational wave spectra due to sound waves only, for different values of $\Lambda$ and $C$. The sensitivity curves of the detectors are peak-integrated sensitivity curves (PISC) \cite{Schmitz:2020syl}.}
    \label{fig:spectrum peaks}
\end{figure}

\begin{figure}
    \centering
    \includegraphics[width=0.35\paperwidth]{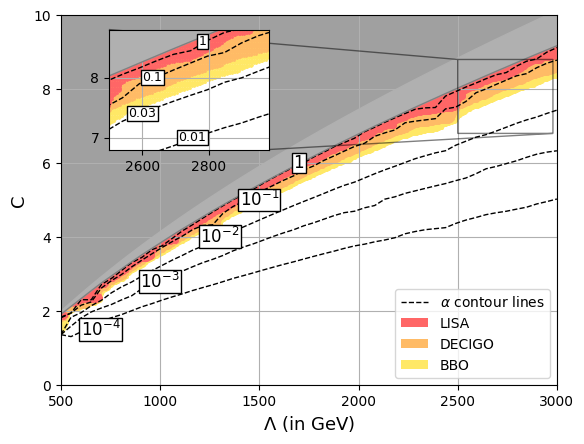}\\    \includegraphics[width=0.35\paperwidth]{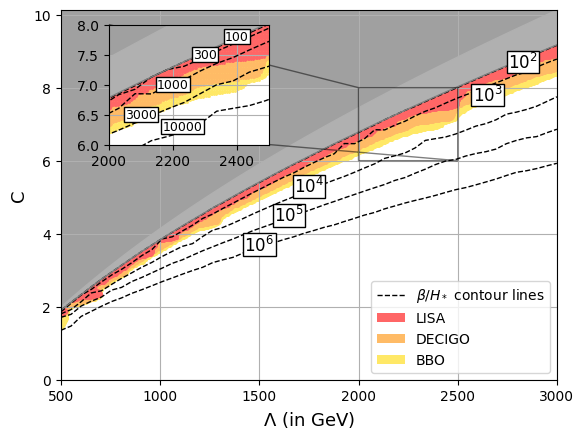}
    \caption{Contour plot for (above) $\alpha$ and (below) $\beta$ in the $(\Lambda, C)$ plane. Colored regions correspond to points within detectability range of different interferometers (obtained by requiring the theoretical amplitude to be larger than the PISC at the peak frequency). The gray regions for larger values of $C$ correspond to the region of metastability (light) and lack of symmetry breaking at $T=0$ (dark gray).}
    \label{fig:ab}
\end{figure}

Peak amplitudes from the sound wave contributions to the GW spectrum are shown in fig.~\ref{fig:spectrum peaks} together with the corresponding PISCs for LISA, DECIGO and BBO. The figure illustrates the behaviour of the spectrum as we vary the parameters $\Lambda$ and $C$. For a fixed $\Lambda$, varying the parameter $C$ by a few $\%$ can change the position of the peak amplitude by 2 orders of magnitude. Conversely, if we double the value of $\Lambda$, the value of $C$ has to be approximately doubled as well in order to keep the peak amplitude at the same position. Thus the spectrum is much more sensitive to $C$ than to $\Lambda$. This can be understood from the following perspective. In this model we are merely altering the shape of the effective potential, leaving all the other sectors as in the SM, so we expect all the new physics to be encapsulated in the scalar field's self-couplings. Looking e.g. at eq.~\eqref{eq:kappa} one sees that the trilinear\footnote{Recall that the trilinear is the first coupling to be modified by the new physics, since we have renormalized the potential such that its first and second derivatives remain unaltered at the minimum.} is much more sensitive to a change in $C$ than to $\Lambda$, as it behaves like $\kappa_\lambda = 1 + \kappa_{\text{1-loop}}^{\text{SM}} + \frac{\text{cnst.}}{\Lambda^2}\left(\frac{C}{1-Cv^2/\Lambda^2}\right)^3$. The new physics contributions to this coupling roughly scales with $\Lambda$ as $\Lambda^{-2}$, whereas an enhancement in $C$ not only boosts this contribution by a factor $C^3$ in the numerator, but it also decreases the (cubed!) denominator, leading to a further enhancement factor.

It is also interesting to investigate the typical values of $\alpha$ and $\beta$ that lead to a detectable GW spectrum, as well as the associated values of $C$ and $\Lambda$. This is shown in figure~\ref{fig:ab}. One notes that BBO could start probing the phase transition for mild values of $\alpha\gtrsim 0.03$ and $\beta/H_* \gtrsim 3000$. However, detectability at LISA requires rather extreme values of $\alpha\sim 1$ and $\beta/H_*\sim \text{few}\times 100$. Such values are within reach of the model under study, but then the parameters would be rather close to the forbidden region of metastability. This means that, if LISA would be able to probe GWs from this model, it would also be able to establish a stringent constrain on $C$~vs.~$\Lambda$.

\begin{figure}
    \centering
    \includegraphics[width=\linewidth]{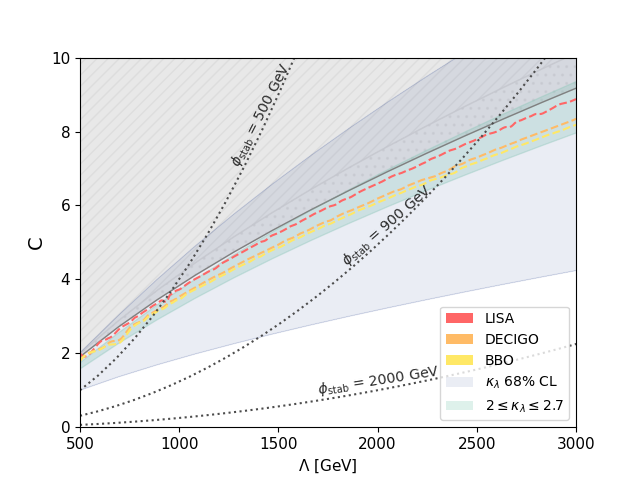}
    \caption{Dashed curves show the values of $C$ and $\Lambda$ for which this model predicts a cosmological gravitational wave background with SNR $=1$ at LISA, DECIGO and BBO. The gray dotted lines represent the values of $\phi_\text{stab}$ at which the potential becomes complex-valued. Also shown are the currently allowed region from 68\% C.L. bounds on the Higgs trilinear coupling. This region is much wider than the parameter range assessible at GW interferometers, which corresponds to $2\leq\kappa_\lambda\leq 2.7$. The hashed regions at larger values of C indicate metastability (dotted) and lack of symmetry breaking at T = 0 (hashed bars). 
    }
    \label{fig:snr plot}
\end{figure}

Our main results are summarized in figure~\ref{fig:snr plot}, illustrating how GW detectors and current/near-future colliders would fare in probing such a logarithmic modification in the scalar effective potential. Again we show the excluded regions (gray hashed) due to metastability of the false vacuum and no symmetry breaking. Also shown are the $\text{SNR}=1$ curves for LISA, DECIGO and BBO. The dotted lines correspond to the regions where the state $\langle \phi\rangle$ becomes unstable at a certain energy level $\phi_\text{stab}$, as discussed in section~\ref{sec:physicality}. Requiring stability up to $2$~TeV excludes most of the interesting region, but a stable state up to $500-1000$~TeV would still allow for a substantial slice of the parameter space. Nevertheless, one sees that this condition clearly establishes an important constraint on the model. Even if this state is not stable, a more detailed analysis would be required to establish its actual decay time, which could still be longer than the age of the Universe and therefore safe. This is however beyond the scope of the current work, and we will deem satisfactory the fact that some regions are still allowed in the $\sim 1$~TeV energy range.

Most interesting in this figure is the wide blue region showing the 68\% C.L. bounds on the Higgs trilinear coupling, $1.1\leq \kappa_\lambda \leq 4.8$~\cite{ATLAS:2022jtk}. Notice how the GW sensitivity curves are well within this region: this means that, should a spectrum be detected and traced to a modification in the scalar potential alone, then these experiments would yield much more precise bounds on the trilinear than current colliders. For comparison we also show that the sensitivity of these GW detectors roughly corresponds to the region of $2.0\leq \kappa_\lambda\leq 2.7$, i.e. an uncertainty $\delta \kappa_\lambda \equiv \kappa_\lambda-1\lesssim 1.7$. This level of uncertainty is expected to be reached only by future lepton colliders operating at $350$~GeV and $200~\text{fb}^{-1}$ luminosity, or a combination of the HL-LHC plus a lepton collider~\cite{DiVita:2017vrr}. We recall, however, that LISA is already a funded project with launch date for the mid 2030s, whereas unfortunately no lepton colliders have reached such an advanced stage yet. Thus, GW detectors could play an important role in tightening current constraints on Higgs self-couplings.

\section{Conclusions}
\label{sec:conclusions}

In this work we have investigated the capabilities of collider experiments and GW detectors in probing deviations of the SM affecting the scalar potential only. For this purpose we have considered the so-called MMFMM (of \emph{minimal modified functional measure model}), a model where this modification stems from a Riemannian definition for the functional integration measure, which would lead to a new logarithmic correction to the Higgs effective potential. However, we expect our results to be qualitatively correct for any model where the BSM physics affects the scalar self-couplings alone.

Our main conclusions are as follows. First, we have seen that there is a strong correlation between the scalar self-couplings (in particular the trilinear) and the condition of detectability of a stochastic GW background produced by a cosmological phase transition. In other words, should the trilinear be sufficiently large (here $\kappa_\lambda \gtrsim 2$), then the GW spectrum shall be detectable at future interferometers such as LISA, DECIGO and BBO ($\text{SNR}>1$). This is not unexpected: if the only new physics is in the scalar sector, then all its effects (including the phase transition strength and the resulting GW spectrum) should be encapsulated in the self-couplings.

Secondly, this means we could use these experiments as a probe of the scalar self-coupling, and it is worth noticing that GW detectors will be probably launched sooner than collider counterparts with similar capabilities. We are entering an era in which GW experiments are receiving increasing attention (and funding), because we have just started probing this spectrum and, being still largely unexplored, it still contains enormous potential of new (potentially groundbreaking) discoveries. It is thus essential that we exploit these experiments with the purpose of probing particle physics as well, rather than relying only on colliders and direct detection. In a recent work we have shown that GW experiments and colliders/detectors could offer complementary results when the new physics lives in a dark sector~\cite{Arcadi:2023lwc}. Here we show instead how \emph{competitive} they can be with each other in probing the scalar sector, assuming everything else remains SM-like.

Our quantitative results are obviously strongly dependent on the particular model we studied. However, we expect them to be qualitatively valid regardless of 
model details,
\emph{as long as only the scalar sector is modified}. Too large self-couplings would lead to stronger transitions and larger GW spectra, until an upper limit is reached where the model becomes unphysical (e.g. failing to lead to a successful symmetry breaking mechanism). The LISA detectability bound is often very close to this upper bound, meaning that either LISA does not see the transition, or, if it does, it also constraints the self-coupling to a very decent accuracy. Other detectors could be sensitive to even lower values of the trilinear, closer to the SM value, therefore leading to its measurement with even better precision.

As a final remark, let us emphasize our earlier statement that \emph{by no means GW detectors will completely obliterate the importance of colliders and direct detection experiments in the near-future}, and that it is crucial to keep investing on all these fronts, especially at this moment when we still know very little about the actual properties of the BSM physics. A detection of a stochastic GW background would allow for a (more or less) direct mapping to the thermodynamical parameters such as $T, \alpha$ and $\beta$~\cite{Gowling:2022pzb, Caprini:2024hue}, but converting these to actual effective couplings depends on certain model assumptions. The same holds for the data collected by colliders, of course, but since they measure different observables (cross sections $\times$ branching ratios) there is more potential for disambiguation if both sets of experiments are used at the same time.

\section*{Acknowledgements}

GFV would like to thank FAPEMIG and Universidade Federal de Minas Gerais (UFMG) for financial support during the preparation of this work.
IK is funded by the National Council for Scientific and Technological Development (CNPq), grant numbers 303283/2022-0 and 401567/2023-0.

\end{document}